\documentstyle[12pt]{article}
\begin{document}
\begin{titlepage}
\rightline{EWHA-TH-003;\quad KHTP-94-02;\quad SNUTP-94-26}
\def\today{\ifcase\month\or
        January\or February\or March\or April\or May\or June\or
        July\or August\or September\or October\or November\or December\fi,
  \number\year}
\rightline{April 1994}
\vskip 1cm
\centerline{\large \bf Thermodynamics of Non-Relativistic Scattering
Theory}
\vskip 1cm
\centerline{{\bf Changrim Ahn}*, {\bf Kong-Ju-Bock Lee}\footnote{
E-mail addresses:\ ahn@benz.kotel.co.kr,\ kjblee@krewhacc.bitnet,\
nam@nms.kyunghee.ac.kr} }
\centerline{{\it Department of Physics} }
\centerline{{\it Ewha Womans University, Seoul, 120-750, Korea} }
\vskip 0.1in
\centerline{{\bf Soonkeon Nam}*}
\centerline {{\it Research Institute for Basic Sciences}}
\centerline {{\it and Department of Physics}}
\centerline {{\it Kyung Hee University, Seoul, 130-701, Korea}}
\vskip .5cm
\centerline{\bf Abstract}
We use thermodynamic Bethe ansatz to study
nonrelativistic scattering theory of low energy excitations of
1D Hubbard model, using the $S$-matrices
proposed by E\ss ler and Korepin.
This model can be described by two types of excitation states, holons and
spinons, as asymptotic states. In the attractive regime,
the spinon is massive while the holon is massless. The situation
is reversed with a repulsive coupling.
We derive that the central charge of the Hubbard model in the IR limit
is $c=1$ while it vanishes in the UV limit.
The contribution is due to the massless degree of freedom, i.e. the holon for
the attractive regime, and the massive mode decouples completely.
This result is consistent with various known results
based on lattice Bethe ansatz computations.
Our results make it possible to use the $S$-matrices of the excitations
to compute more interesting quantities like correlation functions.
\end{titlepage}
\newpage


\def\beq{\begin{equation}}
\def\eeq{\end{equation}}
\def\bea{\begin{eqnarray}}
\def\eea{\end{eqnarray}}
\renewcommand{\arraystretch}{1.5}
\def\ba{\begin{array}}
\def\ea{\end{array}}
\def\bce{\begin{center}}
\def\ece{\end{center}}
\def\nn{\noindent}
\def\pbx{\partial_x}
\def\lbr{\left(}
\def\rbr{\right)}
\def\half{\frac{1}{2}}


\def\ptl{\partial}
\def\al{\alpha}
\def\be{\beta}
\def\ga{\gamma}
\def\Ga{\Gamma}
\def\de{\delta} \def\De{\Delta}
\def\ep{\epsilon}
\def\vep{\varepsilon}
\def\ze{\zeta}
\def\et{\eta}
\def\th{\theta} \def\Th{\Theta}
\def\vth{\vartheta}
\def\io{\iota}
\def\ka{\kappa}
\def\la{\lambda}
\def\La{\Lambda}
\def\rh{\rho}
\def\si{\sigma} \def\Si{\Sigma}
\def\ta{\tau}
\def\up{\upsilon}
\def\Up{\Upsilon}
\def\ph{\phi}
\def\Ph{\Phi}
\def\vph{\varphi}
\def\ch{\chi}
\def\ps{\psi}
\def\Ps{\Psi}
\def\om{\omega}
\def\Om{\Omega}


\def\vol#1{{\bf #1}}
\def\nupha#1{Nucl. Phys. \vol{#1} }
\def\phlta#1{Phys. Lett. \vol{#1} }
\def\phyrv#1{Phys. Rev. \vol{#1} }
\def\PRL#1{Phys. Rev. Lett \vol{#1} }
\def\prs#1{Proc. Roc. Soc. \vol{#1} }
\def\PTP#1{Prog. Theo. Phys. \vol{#1} }
\def\SJNP#1{Sov. J. Nucl. Phys. \vol{#1} }
\def\TMP#1{Theor. Math. Phys. \vol{#1} }
\def\ANNPHY#1{Ann. Phys. (N.Y.) \vol{#1} }
\def\PNAS#1{Proc. Natl. Acad. Sci. USA \vol{#1} }

There has been considerable interest in the interplay of
integrable quantum field theory and statistical mechanics\cite{KorepinBook}.
In particular, a lot of progress in this relationship has been made in
two-dimensional models.
One of the most useful methods in these models
is the factorizable $S$-matrix theory\cite{factS}.
In $1+1$-dimensional integrable field theories where infinite number of
conservation laws exist, the $S$-matrices are facorizable into two-body
elastic $S$-matrices, which satisfy the Yang-Baxter equation.
With known particle spectrum and additional symmetries, one can determine
the $S$-matrices exactly\cite{factS} by solving the Yang-Baxter equation.
Another method  is to diagonalize the Hamiltonian using Bethe ansatz and
find physical particle states and their $S$-matrices\cite{fadeev}.
In addition to their importance as
physical amplitudes of scattering between asymptotic particle states,
exact $S$-matrices can give other interesting quantities like
the central charges of underlying conformal field theories (CFT's)
from the finite size effects, conformal dimensions of the operators,
and even the correlation functions.

Our motivation is to establish the $S$-matrix
approach to study the non-relativistic lattice models like the 1D Hubbard
model\cite{LiebWu}.
Although the Bethe ansatz method is quite useful in finding eigenvalue
spectra, excitation states, and their thermodynamic properties,
it is not so useful in finding other important quantities, in particular,
correlation functions.
Recently, E\ss ler and Korepin derived the $S$-matrices for
low lying excitations of the 1D Hubbard
model\cite{Korepin}.
What concerns us here, as a first step towards the complete
$S$-matrix bootstrap of the low lying excitations of the
Hubbard model, is how to confirm the validity of these matrices.
For this purpose, we employ thermodynamic Bethe ansatz (TBA) for 2D QFT
which is now a standard tool\cite{TBA} to check the $S$-matrices.
In the original formulation of factorizable scattering theory,
one is interested in the scattering of relativistic excitations.
However, in interacting 1-D quantum systems,
such as the Hubbard model, there are in general several
low-energy excitations,
with complicated dispersion relations
and with different Fermi velocities.
Also, the motions of low energy excitations are far from relativistic.
Thus, what we should do is to extend this TBA method to non-relativistic
theories to compute the central charges in the IR and UV limits and
compare with results of finite size effects\cite{Woyn}.

Among the lattice models of the strongly correlated electron systems
in low dimensions, it is believed that the two-dimensional Hubbard model
provides some of the properties of high-$T_{c}$
superconductivity\cite{Anderson}. Furthermore, a strong quantum fluctuation
in low dimensions suggests common features in 2D and 1D Hubbard models.
Fortunately 1D Hubbard model can be exactly
diagonalized via the Bethe-ansatz technique\cite{LiebWu}.
Its thermodynamic properties, such as the susceptibility, the
magnetization, and the low-temperature specific heat, for the repulsive
($U>0$) and the attractive ($U<0$) on-site interactions have been studied in
the literature\cite{ThermoHub}.
It is noticeable that the excitation spectrum is described by the spin and
charge excitations, i.e. spinons and holons,
and the spin (charge) excitation possesses a gap in the
attractive (repulsive) Hubbard model as long as the on-site interaction $U$
exists. The low-energy charge (spin) excitations for the attractive
(repulsive) case are proportional to those for the  antiferromagnetic
Heisenberg chain irrespective
of the strength of the interaction and of the electron filling.
Woynarovich and Eckle\cite{Woyn} have analyzed the finite-size corrections
in the Hubbard model for the repulsive and half-filled case,
and their analysis
yields the central charge of the Virasoro algebra $c=1$ which is
contributed by the spin excitations, while the contribution of the charge
excitations is negligible only if the on-site repulsion $U$ is not so weak.

Let us now examine the dispersion relations of the excitations
in the attractive case.
Note that the results for the repulsive case can be obtained from the
attractive case by interchanging the roles of holons and spinons.
The holon energy in terms of rapidity $\la$ is given by
\beq\label{10}
E_{c}(\la) = 2 \int_{0}^{\infty} \frac{d\om}{\om}
\frac{J_{1}(\om)\cos(\om\la)}{\cosh (\om U)},
\eeq
whereas, momentum is given by
\beq\label{11}
P_{c}(\la)= - \int^{\infty}_{0} \frac{d\om}{\om}
\frac{J_{0}(\om)\sin(\om\la)}{\cosh (\om U)}, \quad
-\frac{\pi}{2}\leq  P_{c}(\la) \leq \frac{\pi}{2}.
\eeq
In the above $U$ is the coupling constant of the on-site interaction of
the 1D Hubbard model.
Notice that the holon is massless and
$E_c\sim v_c P_c$, as $\la\to\infty$, where
$v_c$, the Fermi velocity, is given by
$v_c =E'_c(\infty)/P'_c(\infty)$.

Spinon has the following dispersion relations, for energy,
\beq\label{12}
E_{s}(k) = 2|U|-2\cos k+ 2 \int_{0}^{\infty} \frac{d\om}{\om}
\frac{J_{1}(\om)\cos(\om\sin k)}{\cosh (\om U)}\exp(-|\om U|),
\eeq
and momentum,
\beq\label{13}
P_{s}(k)= k - \int^{\infty}_{0}   \frac{d\om}{\om}
\frac{J_{0}(\om)\sin(\om\sin k)}{\cosh (\om U)}\exp(-|\om U|), \quad
-\frac{\pi}{2}\leq  P_{s}(k) \leq \frac{\pi}{2}.
\eeq
We see that the spinon stays massive for finite $U$ for all values
of the rapidity.

Recently, using these dispersion relations, E\ss ler and Korepin derived
the scattering matrices of Hubbard model from the Bethe ansatz solution,
by generalizing a method of extracting $S$-matrix from the asymptotics
of the wave functions of the scattering state\cite{Korepin}.
These $S$-matrices of the excitations on the lattice are well-defined
as far as the wave packets of two excitations are well-separated.
The complete scattering matrix is $16\times 16$
dimensional and is in a block diagonal form consisting of 4 blocks.
Each of the blocks describe scatterings of
holon-holon, spinon-holon, holon-spinon, and spinon-spinon, respectively,
as follows:
\beq \label{21}
S=\left(\begin{array}{cccc}
S_{cc}(u) & 0 & 0 & 0 \\
0 & S_{sc}(w) & 0 & 0 \\
0 & 0 & S_{cs}(w) & 0 \\
0 & 0 & 0 & S_{ss}(v)
\end{array}\right), \quad
\eeq
where the holon-holon scattering amplitude is,
\beq \label{22}
S_{cc}(u) =
-\frac{\Ga \lbr\frac{1+iu}{2}\rbr \Ga\lbr 1-\frac{iu}{2} \rbr}
{\Ga \lbr\frac{1-iu}{2}\rbr \Ga\lbr 1+\frac{iu}{2} \rbr}
\lbr \frac{u}{u+i} {\bf I} + \frac{i}{u+i} {\bf P} \rbr,
\quad u=\frac{\la-\la'}{2|U|},
\eeq
with $\Ga$ being the gamma function.
In the above $\bf I$ and $\bf P$ are the $4 \times 4$ identity and permutation
matrices respectively.
This $S$-matrix has the same form as that of the spin $\frac{1}{2}$
Heisenberg antiferromagnet and of the $SU(2)_1$ WZNW model\cite{Wiegmann}.
The spinon-spinon scattering amplitude is
\beq \label{23}
S_{ss}(v) =
\frac{\Ga \lbr\frac{1-iv}{2}\rbr \Ga\lbr 1+\frac{iv}{2} \rbr}
{\Ga \lbr\frac{1+iv}{2}\rbr \Ga\lbr 1-\frac{iv}{2} \rbr}
\lbr \frac{v}{v-i} {\bf I} - \frac{i}{v-i} {\bf P} \rbr,
\quad v=\frac{\sin k- \sin k'}{2|U|},
\eeq
and can be obtained from $S_{cc}$ by setting $u\rightarrow -v$.
The spinon-holon scattering amplitude is,
\beq \label{24}
S_{sc}(w) =  -i\frac{1+i\exp (\pi w)}{1-i\exp (\pi w)}{\bf I}, \quad \quad
w=\frac{\la-\sin k}{2|U|}.
\eeq
We have the same form for holon-spinon scattering amplitude $S_{cs}$.
Notice that $S_{sc}$ and $S_{cs}$ approach constant values as
$w \rightarrow \infty$.

Let us now formulate the TBA for non-relativistic scattering.
The TBA computes the Casimir energy of a theory on a circle of length
$R$ with $S$-matrices and particle spectrum as input data\cite{TBA}.
With a temperature $T=1/R$ the configuration of minimizing free energy
gives the ground state energy of the system, which is again related
to the central charges of the underlying UV CFT by
\beq
E_{\rm ground}(R)\sim -\sum_{i}\frac{\pi c_{i}}{6v_{i}R},
\eeq
where $v_{i}$ are Fermi velocities for excitations in the system and
$c_{i}$ are the corresponding effective central charges.

Consider $N$ particles in a box of length $L$ with periodic boundary
condition(PBC).
Moving a $k$-th particle of type $a$ with energy $E_{a}(\th_{k})$
and momentum $P_{a}(\th_{k})$ all the way by exchanging with other particles
and coming back to the original configuration using PBC, we get
\beq\label{pbc}
e^{-iLP_{a}(\th_{k})}\prod_{i=1, i\neq k}^{N}S_{aa_{i}}(\th_{k}-\th_{i})=1,
\eeq
where the index $a_{i}$ specifies species of the $i$-th particle.
In general the product of $S$-matrices is a large size matrix called
transfer matrix and one should diagonalize this by some techniques.
However, we consider a diagonal scattering theory first because the
non-diagonal case can be understood as a slight modification.
Taking logarithms on both sides of Eq.(10) gives
\beq
-LP_{a}(\th_{k})+\sum^{N}_{i=1}\frac{1}{i}\ln S_{aa_{i}}(\th_{k}-\th_{i})
=2\pi n_{k} ,
\eeq
with an arbitrary integer $n_{k}$. In the thermodynamic limit, $N$,
$L\rightarrow \infty$, one can express Eq.(11) in terms of density of
states like
\beq
2 \pi \rh_{a}(\th) = -L P'_{a}(\th)
+ \sum _{b}\int d\th'\rh_b^1(\th')\ph_{ab}(\th-\th'),
\eeq
where $\rho_{a}(\th)$ and $\rho^{1}_{a}(\th)$ are the densities of allowed
and occupied states, respectively, and $\ph_{ab}$ is the logarithmic
derivatives of $S$-matrices $S_{ab}$. In terms of `pseudo-energies'
$\ep_{a}$ defined
by ${\rm e}^{-\ep_{a}}=\rh^{1}_{a}/(\rh_{a}-\rh^{1}_{a})$, one can express
the ground state energy by
\beq\label{104}
E_{\rm ground}(R)=-{1\over{R}}\sum_a\int_{-\infty}^{\infty}{d\th\over{2\pi}}
RP'_a(\th)L_{\ep_a}(\th),
\eeq
where $L_{\ep}(\th)=\ln[1+{\rm e}^{-\ep(\th)}]$.
$\ep_{a}$ is determined by the minimizing condition of the free
energy which is the following set of nonlinear equations:
\beq
\ep_a(\th)=RE_a(\th)-\sum_b\ph_{ab}*L_{\ep_b}(\th),
\eeq
where $*$ denotes rapidity convolution,
$f*g(\th)=\int_{-\infty}^{\infty}d\th' f(\th-\th')g(\th')$.
As we mentioned above, the sum in Eq.(13) for non-diagonal theories
has to been taken with care. Diagonalization of transfer matrix
brings in additional `massless' particles, which do not contribute
to the central charge directly in the sum, due to masslessness, but
which nevertheless affect the massive particle distributions.
Additional care for non-relativistic scattering is that
the participating particles can have different Fermi velocities,
unlike in the relativistic case where all have the light speed as
Fermi velocity.

For the non-diagonal theories, the product of $S$-matrices in Eq.(10) is
replaced by the eigenvalues of the transfer matrix.
For 1D Hubbard model the nondiagonal matrices $S_{cc}$ and $S_{ss}$ are
six vertex model type:
\beq
S_{\al} = \left(\begin{array}{cccc}
a_{\al} & 0 & 0 & 0 \\
0 & b_{\al} & c_{\al} & 0 \\
0 & c_{\al} & b_{\al} & 0 \\
0 & 0 & 0 & a_{\al}
\end{array}\right), \quad \al= c,s. \quad
\eeq
The eigenvalues of the transfer matrices and the associated constraint
equations can be derived by algebraic Bethe Ansatz method to be
\beq
\La_{\al}(\th)=\prod_{i=1}^{N}a_{\al}(\th-\th_i)
\prod_{r=1}^{M}{a_{\al}(y_r-\th)\over{
b_{\al}(y_r-\th)}}+\prod_{i=1}^{N}b_{\al}(\th-\th_i)\prod_{r=1}^{M}
{a_{\al}(\th-y_r)\over{b_{\al}(\th-y_r)}},
\eeq
and
\beq
\prod_{i=1}^{N}{b_{\al}(y_k-\th_i)\over{a_{\al}(y_k-\th_i)}}\prod_{r=1}^{M}
{b_{\al}(y_r-y_k)\over{a_{\al}(y_r-y_k)}}
{a_{\al}(y_k-y_r)\over{b_{\al}(y_k-y_r)}}=-1.
\eeq
{}From the $S$-matrices Eqs.(6) and (7), one can read off the corresponding
matrix elements to evaluate the explicit eigenvalues.
The holon and the spinon sectors are coupled by diagonal scattering matrix
$S_{sc}$.

Hence we have two kinds of the periodic conditions for
the Bethe wave functions
of holons and spinons in terms of $\La_c$, $\La_s$, and $S_{cs}$
and two sets of the constraint equations.
For simplicity, we concentrate on holon sector first and will
generalize the argument to spinons.
{}From Eqs.(6) and (17) the constraint equation for holon sector becomes
\beq
\prod_{i=1}^{N}\frac{\be_{k}-\th_{i}-i|U|}{\be_{k}-\th_{i}+i|U|}
\prod_{r=1}^{M}\frac{\be_{k}-\be_{r}+2i|U|}{\be_{k}-\be_{r}-2i|U|} = -1,
\eeq
where we have introduced the shifted rapidities $\beta_{i}=y_{i}-i|U|$
to have the unitary form.
It is well known, from the analogy in the antiferromagnetic Heisenberg chain,
that in the thermodynamic limit, $N \rightarrow \infty$,
the general solutions
of these equations are the strings consisting of $n$-pseudoparticles of roots
$\be_{r}^{n,j}= \be_{r}^{0}+i|U|(n+1-2j)$, where $\be_{r}^{0}$ is real,
$j=1, \cdots, n$, and $n=1,\cdots, \infty$.
Such a string is a bound state of $n$-pseudoparticles
and can be interpreted as a fictitious massless particle of
real rapidity $\be_{r}^{0}$.
Since the length of the strings could be infinitely long,
there are infinite number of such massless particles.
Similarly the constraint equations for spinons can be
understood in the context of another kind of pseudoparticles designated
by the rapidities $\al$'s which also form the string-solutions in the
thermodynamic limit.
Therefore what we have is a diagonal scattering theory of holons, spinons
and infinite number of massless particles associated with them.
The scattering amplitudes of $n$-th massless particle with
holon is
\beq
S_n(\th)=\frac{\th-in|U|}{\th+in|U|},
\eeq
and the scattering amplitudes between the massless particles are
\bea
S_{nm}(\th)=\left[ \frac{\th+i|n-m||U|}{\th-i|n-m|{|U|}} \right]
\qquad\qquad\qquad\qquad\qquad\qquad\qquad\qquad\qquad\qquad
\nonumber \\
\times\left[ \frac{\th+i(|n-m|+2){|U|}}{\th-i(|n-m|+2){|U|}}\cdots
   \frac{\th+i(n+m-2){|U|}}{\th-i(n+m-2){|U|}}\right]^{2}
   \left[ \frac{\th+i(n+m){|U|}}{\th-i(n+m){|U|}}\right].
\eea
For the spinons, the corresponding scattering amplitudes are
obtained by the replacement $\th\to -\th$.

The minimizing condition of the free energy using the above mentioned
$S$-matrices leads to an infinite set of non-linearly coupled equations.
It is a standard procedure to use Fourier transformation
on the TBA equations
and to simplify the equations in terms of a unified kernel
$\varphi=\left(4|U|\cosh \frac{\pi \th}{2|U|}\right)^{-1}$\cite{ZamApp}:
\bea
RE_{c}(\th)&=&\ep_{0}(\th)+\varphi*(L_{\ep_{1}}+L_{\et_{0}})(\th)
\nonumber\\
0&=&\ep_{n}(\th)+\varphi * (L_{\ep_{n-1}}+L_{\ep_{n+1}})(\th),
\quad n\geq 1\\
RE_{s}(\th)&=&\et_{0}(\th)+\varphi*(L_{\et_{1}} +L_{\ep_{0}})(\th)
\nonumber\\
0&=& \et_{n}(\th)+\varphi * (L_{\et_{n-1}}+L_{\et_{n+1}})(\th),
\quad n\geq 1.\nonumber
\eea
These TBA equations have the incidence structure of an infinite chain,
where a pair of semi-infinite chains of SU(2) invariant factorized
scatterings are joined together as shown in the following picture:
\vskip 30pt
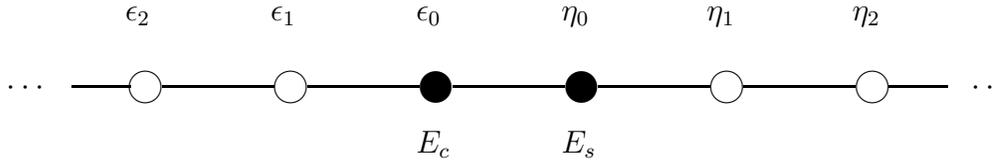
\begin{picture}(400,50)(0,-25)
\put (-5,-3){$\cdots$}
\put (20,0){\line(10,0){22}}
\put (47.5,0){\circle{12}}
\put (54,0){\line(10,0){42}}
\put (102.5,0){\circle{12}}
\put (109,0){\line(10,0){42}}
\put (157.5,0){\circle*{12}}
\put (164,0){\line(10,0){42}}
\put (212.5,0){\circle*{12}}
\put (219,0){\line(10,0){42}}
\put (267.5,0){\circle{12}}
\put (274,0){\line(10,0){42}}
\put (322.5,0){\circle{12}}
\put (329,0){\line(10,0){22}}
\put (360,-3){$\cdots$}
\put (150,-25){$E_c$}
\put (205,-25){$E_s$}
\put (150,25){$\epsilon_0$}
\put (205,25){$\eta_0$}
\put (95,25){$\epsilon_1$}
\put (260,25){$\eta_1$}
\put (40,25){$\epsilon_2$}
\put (315,25){$\eta_2$}
\end{picture}

While the TBA equations are non-linearly coupled equations and are hard
to solve explicitly, it is rather easy for the cases of
UV($R\rightarrow 0$) and IR ($R\rightarrow \infty$) limits.
However, the analysis of these two limits depends on the energy-momentum
dispersion relations.

(i) Contrary to the relativistic case where the central charge
can still get nontrivial contributions in the UV limit because
$E(\th)$ can be large enough to overcome vanishing $R$ in Eq.(13)
at large rapidity, the central charge becomes zero for the
Hubbard model as $R\rightarrow 0$ since
$P'_{c}(\th)$ and $P'_{s}(\th)$ are bound over all values of rapidities.

(ii) The situation changes in the IR limit.
When $U$ is finite, the spinons become massive and
do not contribute to the central charge because the pseudo-energy
$\et_0(\th) \sim  R E_s(\th)\rightarrow \infty$ for all values of $\th$.
After decoupling spinon sector, we have only semi-infinite chain of holon
sector.
For a finite value of $\th$, $E_c(\th)$ is non-zero and from Eq.(21),
$\ep_0(\th)$ becomes infinite.
Therefore, only non-vanishing contribution to the central charge comes from
$\th\to\infty$ limit.
In this limit, taking a derivative on Eq.(14) and substituting $P'_c$ with
$E'_c/v_c$ into Eq.(13),
we can now invoke the standard TBA analysis to evaluate the central charge
in terms of the pseudo-energies at $\th=0$ and $\infty$. Note that
the Fermi velocity in the denominator of Eq.(9) is canceled by the $v_c$
in the above substitution.
The final result for the central charges is
\beq
c_{\rm eff} = {6\over{\pi^{2}}}\sum_{n=0}^{\infty}\left[
{\cal L}\left({x^{\infty}_{n}\over{1+x^{\infty}_{n}}}\right)
-{\cal L}\left({x^{0}_{n}\over{1+x^{0}_{n}}}\right)\right],
\eeq
where ${\cal L}(x)$ is the Rogers dilogarithmic function;
\beq
{\cal L}(x)= -\frac{1}{2}\int^{x}_{o} dt \left[\frac{\ln (1-t)}{t}
+ \frac{\ln t}{(1-t)}\right],
\eeq
and we have defined $x_n^{\infty}=\exp[-\ep_n(\infty)]$ and
$x_n^{0}=\exp[-\ep_n(0)]$. The TBA equations, Eq.(21), can be rewritten
as a set of algebraic equations for $x_n$'s;
\bea
x_{0}^{\infty}=(1+x_{1}^{\infty})^{\half},\quad
x_{n}^{\infty}&=&(1+x_{n-1}^{\infty})^{\half}(1+x_{n+1}^{\infty})^{\half},
\quad n\ge 1,\\
\nonumber
x_{0}^{0}=0,\qquad\qquad\quad
x_{n}^{0}&=&(1+x_{n-1}^{0})^{\half}(1+x_{n+1}^{0})^{\half},\quad n\ge 1.
\eea
These have solutions $x_{n}^{\infty}=(n+2)^{2}-1$ and
$x_{n}^{0}=(n+1)^{2}-1$.
Since $x_{n}^{0}=x_{n-1}^{\infty}$, only $x_{\infty}^{\infty}$ survives
to give $c_{\rm eff}=1$ after using ${\cal L} (1) =\frac{\pi^{2}}{6}$.

As we claimed in the beginning, the central charge we computed comes
from the holon sector while the spinon sector decouples.
This is what happens for finite $U$ but seems valid even vanishing $U$
as far as $U>{1/R}$.
In the literature, a discontinuity between $U\to 0$ and $U=0$ has been
predicted such that if $U=0$ the central charge will be $2$ because
the model is nothing but a theory with four free fermions.
In our analysis, we use the holon and spinon $S$-matrices which are
valid for non-vanishing $U$.
Within this validity, our result is what has been noticed in the
literature from different computations.
With this confirmation for the E\ss ler-Korepin $S$-matrices,
we established the $S$-matrix program for the non-relativistic models.
The most noticeable difference of the non-relativistic theories
from the relativistic ones is that
the IR limit of the former models corresponds to the UV limit of the latter.
This is the limit where massless degree of freedom survives and gives
correct central charges.
We hope our result can be a starting point for the application of
the $S$-matrices to various lattice problems.
What we are particularly interested in is to compute correlation functions
using the form-factor approach\cite{formfactor}.
In this scenario, correlation functions of any local operator are
expressed in terms of the form-factors which can be computed from the
exact $S$-matrices.
Although one needs to sum an infinite terms, this can be realized as
the sum converges very fast.
We hope to report this result elsewhere.

\newpage
\vskip 1cm
\baselineskip 0.6cm
\leftline{\bf Acknowledgement}
We thank Choonkyu Lee and Sungkil Yang for very helpful remarks.
The work of C.A. and K.J.B.L. has been supported in part by
KOSEF 941-0200-003-2.
The work of S.N. is supported in part by non-directed research fund,
Korea Research Foundation, 1993,
and by Ministry of Education (RIBS).
We also acknowledge a partial support from CTP/SNU.

\end{document}